\documentclass[12pt]{article}
\usepackage{graphicx}

\newsavebox{\sboxpubnumber}
\newsavebox{\sboxpubdate}
\newcommand{\pubdate}[1]{\begin{lrbox}{\sboxpubdate}{#1}\end{lrbox}}

\newcommand{\Title}[1]{\begin{center} {\Large #1 } \end{center}}
\newcommand{\Author}[1]{\begin{center}{ \sc #1} \end{center}}
\newcommand{\Address}[1]{\begin{center}{ \it #1} \end{center}}

\newenvironment{Abstract}{\begin{quotation}  }{\end{quotation}}
\newenvironment{Presented}{\begin{quotation} \begin{center}
             PRESENTED AT\end{center}\bigskip
      \begin{center}\begin{large}}{\end{large}\end{center}
      \end{quotation}}
\newcommand{\Acknowledgements}{\bigskip  \bigskip \begin{center} \begin{large}
             \bf ACKNOWLEDGEMENTS \end{large}\end{center}}

\begin{document}

\begin{titlepage}
\pubdate{\today}                    

\vfill
\Title{Environmentally Friendly Renormalization Group
       and Phase Transitions}
\vfill
\Author{F. Astorga\footnote{Funding from Conacyt under grant 32399-E, 
                            SNI and CIC-UMICH is gratefully acknowledged}}
\Address{Instituto de F\'{\i}sica y Matem\'{a}ticas \\ 
         Universidad Michoacana de San Nicol\'{a}s de Hidalgo \\
         Edificio C3, Ciudad Universitaria, Morelia, Michoac\'{a}n, 
         C.P. 58060, M\'{e}xico}
\vfill
\vfill
\begin{Abstract}
       We discuss an environmentally friendly renormalization group 
       approach to analyze phase transitions. We intend to 
       apply this method to the Electroweak Phase Transition.   
       This work is in progress. We present some 
       previously obtained results concerning a $\lambda\phi^{4}$
       theory, where the main features of this algorithm are 
       introduced.


\end{Abstract}
\vfill
\begin{Presented}
    COSMO-01 \\
    Rovaniemi, Finland, \\
    August 29 -- September 4, 2001
\end{Presented}
\vfill
\end{titlepage}
\def\thefootnote{\fnsymbol{footnote}}
\setcounter{footnote}{0}


\section{Introduction}

The actual mechanism of baryogenesis is still elusive. Although 
several scenarios which generate an excess of matter over antimatter
have been proposed, we still lack evidence which 
favours any of them. However, constraints on inputs of the models 
have started ruling out some of these. This seems to be the case 
for the Electroweak Phase Transition (EWPT) within the Standard 
Model where a strong enough first 
order transition is not achieved for realistic Higgs masses~\cite{
Laine}. Even in the supersymmetric extension, the parameter space 
for the desired transition is getting reduced. 

Lattice calculations 
have played an important role in reaching these conclusions and 
even though different approaches have been used to study the 
EWPT, we believe there is still some physical insight to gain 
from analyzing this transition by means of an environmentally 
friendly renormalization group running with the temperature. 
In what follows, we will describe the main steps in the application 
of this method to $\lambda\phi^{4}$. The details can be found 
somewherelse~\cite{Chris}.

\section{Analysis}

The Renormalization Group Equation can be seen as a simple 
consequence of the fact that the physics is independent of 
the arbitrary renormalization scale $\mu$ at which we choose 
to define our parameters

   \[\mu\frac{d\Gamma^{(N)}}{d\mu}=0\]

An environmentally friendly renormalization group realises 
that the effective degrees of freedom of a system may change 
qualitatively during its evolution. 
Such is the case, for instance, of the confinement-
deconfinement transition where we have the change 
between quark-gluon degrees of freedom and hadron-
meson degrees of freedom. Hence, a suitable choice 
of parameters to describe this evolution is convenient, 
otherwise we end up with a description which does not 
fit the new degrees of freedom. In the case of 
finite temperature field theory, a renormalization 
group running with the temperature seems to be 
an adequate method of tracking the evolution of 
the system. 

For $\lambda\phi^{4}$, the Euclidean action is written as follows  
in terms of bare quantities and the inverse temperature 
$\beta=1/T$  

\[S[\phi_{B}]=\int_{0}^{\beta}dt\int d^{d-1}x[\frac{1}{2} 
(\nabla\phi_{B})^{2} + \frac{1}{2} M^{2}_{B} \phi^{2}_{B} 
+ \frac{\lambda_{B}}{4}\phi^{4}_{B}] \]

The renormalized parameters are specified by the  
normalization conditions at an arbitrary temperature 
scale $\tau$ 

\[ \frac{\partial}{\partial p^{2}} \Gamma^{(2)}_ {t}
(p, \bar{\phi_{H}}(\tau), M(\tau), \lambda(\tau), T=\tau)
\mid_{p=0} \; = 1 \]
 \[\Gamma^{(2)}
(p=0, \bar{\phi_{H}}(\tau), M(\tau), \lambda(\tau), T=\tau)
 = M^{2}(\tau) \]
\[\Gamma^{(4)}_ {t}
(p=0, \bar{\phi_{H}}(\tau), M(\tau), \lambda(\tau), T=\tau)
 = \lambda(\tau) \]

where $\bar{\phi_{H}}$ is the renormalized field corresponding 
to a reference external current $H$.

From here, the flow equations describing the running 
of the mass and the coupling are 

\[ \tau \frac{d M^{2}(\tau)}{d \tau} = \beta_{M}, \;\;\; 
\tau \frac{d \lambda(\tau)}{d \tau} = \beta_{\lambda} \]

where the $\beta$-functions $\beta_{M}$ and $\beta_{\lambda}$ 
are to be calculated at each order of approximation. To one loop 
and taking the external current $H=0$ these take 
the following form

\[ \beta_{M}= \left \{\begin{array}{ll} 
\frac{\lambda}{2}\tau \frac{\partial}
{\partial \tau} \bigcirc_{1} &  \tau > T_{c} \\
-\lambda (\tau\frac{\partial}{\partial \tau} \bigcirc_{1} 
+ \frac {3}{2} M^{2} \tau \frac{\partial}
{\partial \tau} \bigcirc_{2}  &  \tau < T_{c}  
\end{array} 
\right. \]

\[ \beta_{\lambda} =  -\frac{3}{2} \lambda^{2} \tau 
\frac{d}{d \tau}\bigcirc_{2}  \]

In this expression, $\bigcirc_{n}$ corresponds to the 
one loop diagram with $n$ propagators, without vertex 
factors, and at zero external momentum. 

The flow equation for the coupling $\lambda$ can be solved 
analytically and substituted into the equation for the mass 
$M$, which needs to be solved numerically. Once this is done, 
the critical temperature $T_{c}$ separating the two phases 
is obtained. Further discussion of this case and comparison of this 
approach and the work of others can be found in~\cite{Chris}. 
The techniques of environmentally friendly renormalization 
can be revised in~\cite{Chris2}.


\section{Conclusions}

Environmentally friendly renormalization groups seem to be 
a fine probe to analyze the electroweak phase transition, 
better equiped than other approaches, though some physical 
intuition of the dynamics is needed. We hope to gain some 
more insight into the details of the EWPT using this method. 
This will add to the knowledge of the phase transition we 
have up to now.

\Acknowledgements

The analysis of the electroweak phase transition by means of 
an enviromentally friendly renormalization group is currently 
being carried out in collaboration with Chris Stephens, 
Axel Weber and Carlos Mendoza.

\end{document}